\documentclass[aps,pop,twocolumn,fleqn]{revtex4}  

\usepackage{amsmath}
\usepackage{amssymb}
\usepackage{graphicx}  

\begin{document}

\title[]{Energy boost in laser wakefield accelerators using sharp density transitions}

\author{A. D\"opp}
\address{LOA, ENSTA ParisTech - CNRS - \'Ecole Polytechnique - Universit\'e Paris-Saclay, 828 Boulevard des Mar\'echaux, 91762 Palaiseau Cedex, France }
\author{E. Guillaume}
\address{LOA, ENSTA ParisTech - CNRS - \'Ecole Polytechnique - Universit\'e Paris-Saclay, 828 Boulevard des Mar\'echaux, 91762 Palaiseau Cedex, France }
\author{C. Thaury}
\address{LOA, ENSTA ParisTech - CNRS - \'Ecole Polytechnique - Universit\'e Paris-Saclay, 828 Boulevard des Mar\'echaux, 91762 Palaiseau Cedex, France }
\author{A. Lifschitz}
\address{LOA, ENSTA ParisTech - CNRS - \'Ecole Polytechnique - Universit\'e Paris-Saclay, 828 Boulevard des Mar\'echaux, 91762 Palaiseau Cedex, France }
\author{K. Ta Phuoc}
\address{LOA, ENSTA ParisTech - CNRS - \'Ecole Polytechnique - Universit\'e Paris-Saclay, 828 Boulevard des Mar\'echaux, 91762 Palaiseau Cedex, France }
\author{V. Malka}
\address{LOA, ENSTA ParisTech - CNRS - \'Ecole Polytechnique - Universit\'e Paris-Saclay, 828 Boulevard des Mar\'echaux, 91762 Palaiseau Cedex, France }

\begin{abstract}
The energy gain in laser wakefield accelerators is limited by dephasing between the driving laser pulse and the highly relativistic electrons in its wake. Since this phase depends on both the driver and the cavity length, the effects of dephasing can be mitigated with appropriate tailoring of the plasma density along propagation. Preceding studies have discussed the prospects of continuous phase-locking in the linear wakefield regime. However, most experiments are performed in the highly non-linear regime and rely on self-guiding of the laser pulse. Due to the complexity of the driver evolution in this regime it is much more difficult to achieve phase locking. As an alternative we study the scenario of rapid rephasing in sharp density transitions, as was recently demonstrated experimentally. Starting from a phenomenological model we deduce expressions for the electron energy gain in such density profiles. The results are in accordance with particle-in-cell simulations and we present gain estimations for single and multiple stages of rephasing.
\end{abstract}

\maketitle

\section{Introduction}

Plasma-based particle accelerators \cite{Esarey:1996ui} use charge separation between electrons and ions to create electric fields in the order of gigavolt to teravolt per meter, exceeding the breakdown-limited field strength in conventional accelerators by several orders of magnitude. The charge separation is induced by a driver, which is either a bright particle or laser beam. The former is known as beam-driven plasma wakefield acceleration (PWFA), which could notably demonstrated energy doubling of a 42 GeV-class electrons beam \cite{Blumenfeld:2007ja} as well as positron acceleration \cite{Corde:2015ii}. The latter is referred to as laser wakefield acceleration (LWFA) and during the past two decades this method has been used to accelerated electrons from rest to first some tens of MeV \cite{Modena:1995wb}, to always higher energies \cite{Malka:2002eu}, reaching GeV-scale \cite{Leemans:2006ux} and recently multi-GeV energies \cite{Wang:2013el,Kim:2013dc,Leemans:2014kp}.

Plasma wakefield acceleration can be seen as a special type of resonance acceleration, whose accelerating structure is a plasma wave. In this kind of accelerator a particle first goes through an injection process, whose primary challenge is to make the particles co-propagate with the wave. While PWFA experiments usually inject part of the drive beam into the wakefield, LWFA almost exclusively relies on the injection of background plasma electrons. Such injection can be facilitated using for instance density downramps \cite{Bulanov1998,Suk:2001uo}, colliding pulses \cite{Esarey:1997wc} or delayed tunneling ionization \cite{Singh:2006bs}.

Once injected the interaction length is essentially limited to the length over which the wave structure can be sustained. While this condition assures that energy exchange between the wave and particles is possible, it is also important to assure that particles interact with the accelerating part of the field, ideally maintaining a synchronous phase $\phi$ with the strongest possible field gradients. Initially this is the case in most injection scenarios, as particles are usually trapped at the very back of the wakefield.
However, for such phase matching the particle velocity $v_e$ and the phase velocity of the wave $v_{\phi}$ need to be the same, which is not necessarily the case. 

In beam-driven plasma wakefield accelerators both driver and witness are highly relativistic, i.e. the associated Lorentz factor $\gamma\gg1$. Both are therefore moving at a velocity close to the speed of light in vacuum $c_0$ (approximately $v_e/c_0\approx 1-1/2\gamma^2$) and the dephasing is not a pressing issue. In travelling wave RF accelerators the phase velocity is in general superluminal, and the synchronization problem is often resolved by disk-loading the cavities, which reduces the phase velocity \cite{wangler2008rf}. However, in laser-driven wakefield acceleration dephasing remains the mayor limitation of achievable energy gain. Here the phase velocity of the plasma wave is of the order of the group velocity of the laser driver, which for a cold underdense plasma is
\begin{equation}  \frac{v_g}{c_0}\approx  1-\frac12\frac{n_e}{n_c}, \end{equation}
where $n_e/n_c\ll1$ denotes the ratio of electron density $n_e$ with respect to the critical density at a laser wavelength $\lambda_0$ ($n_c\approx \lambda_0^{-2}[\mu \mbox{m}]\times 1.1\times 10^{21}$cm$^{-3}$).
Since the group velocity increases at lower electron densities, dephasing is often avoided by reducing the plasma density. Yet this approach has a number number of drawbacks. For example it goes in hand with a reduction of the accelerating field gradient, thus increasing the accelerator length. Furthermore it is harder to self-guide the laser \cite{Gonsalves:2015cc} since the critical power $P_c\sim (n_c/n_e)\times$17 GW \cite{Sprangle:1987bw}. It is therefore of interest to find alternatives to mitigate the effects of dephasing \cite{Katsouleas:1986gv}. 
Such rephasing was recently demonstrated experimentally \cite{Guillaume:2015di} and the aim of this article is to discuss the technique from a theoretical point of view. The paper is structured as follows: First we discuss energy gain in the self-guided blowout regime. We then discuss the problem of phase-locking in this regime and introduce the concept of phase reset. Our analytical estimations are then compared to particle-in-cell (PIC) simulations. We conclude with perspectives for multiple stages of rephasing.

\section{The self-guided highly non-linear blowout regime}

\begin {figure}
\centering
\includegraphics[width=0.96\linewidth]{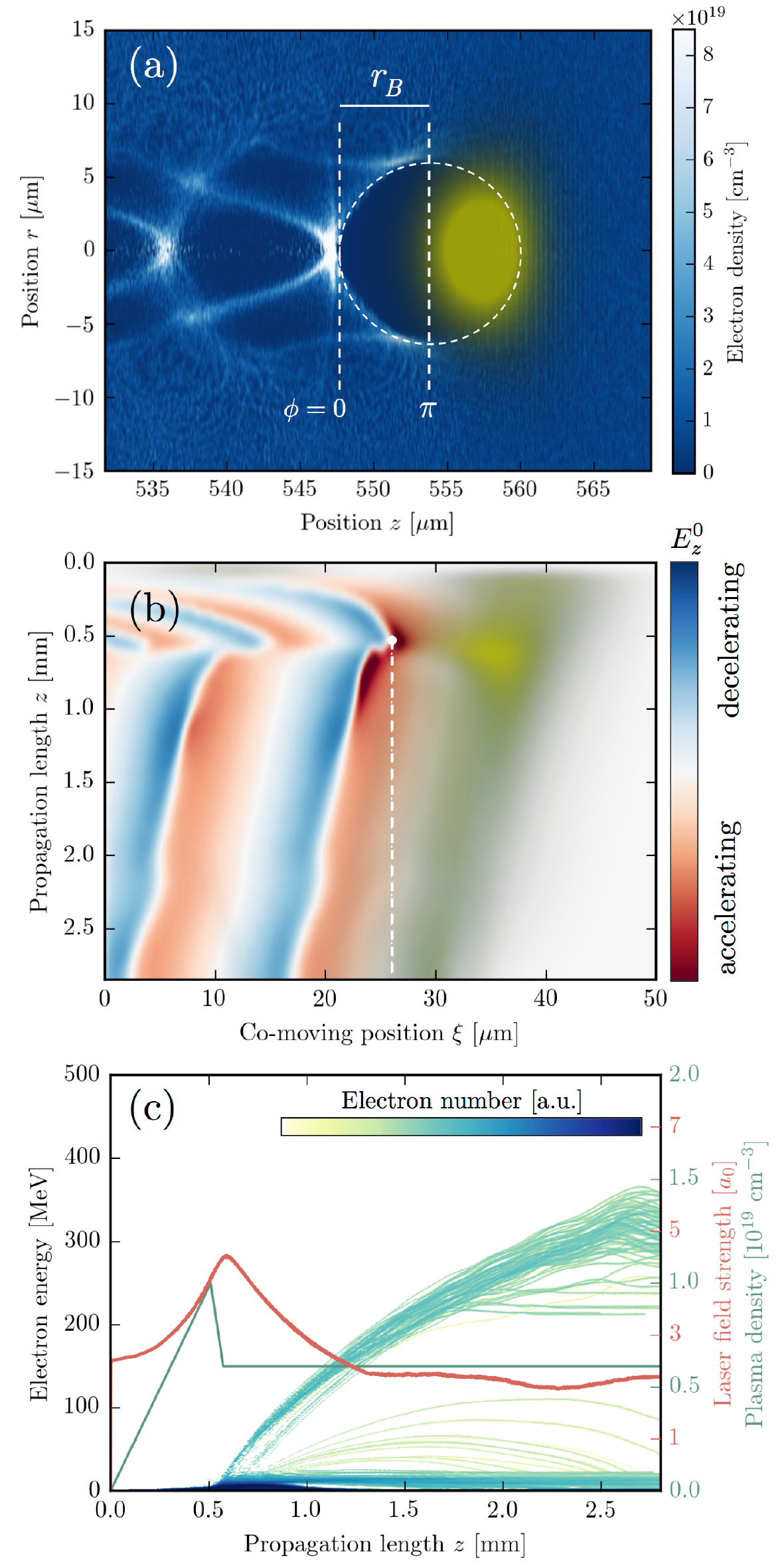}
\caption{Simulation of a laser wakefield accelerator in the blowout regime with injection in a density transition. (a) shows the bubble-shaped ion-cavity (dark blue) that forms behind the laser (yellow). (b) presents lineouts of the on-axis fields $E_z(r=0)$ with accelerating fields in red and decelerating fields in blue. The dashed white line represents the electron bunch, with the dot marking the injection position. (c) shows the plasma density profile, laser pulse evolution and electron acceleration.}
\label{fig1}
\end{figure}

Most laser wakefield accelerators rely on self-focusing to guide the laser driver over distances superior to the Rayleigh length. The laser pulse then often focuses and self-compresses, typically reaching normalized peak field amplitudes in the order of $a_0\sim 4-10$, \cite{Corde:2013gj}. For such intense drivers, the ponderomotive force pushes basically all electrons away, leaving behind a pure ion cavity \cite{Mora:1996jy}. Due to its round shape this cavity is often referred to as \textit{bubble} \cite{Pukhov2002}, which is exemplarily shown in figure \ref{fig1}(a). The cavity scales with the plasma density and the laser intensity. For self-guided laser pulses it was found empirically that the laser focuses to a matched spot size that follows approximately $k_pw_0\simeq 2\sqrt{a_0}$, \cite{Lu:2007eb}.\\
For a perfect circular blowout the potentials inside the bubble have the form
\begin{equation}
\Phi=\frac{k_p^2}{4}\left( r_B^2-r^2\right)
\end{equation}
and it immediately follows that the associated longitudinal fields are linear 
\begin{equation} 
E_z = -\frac{m_e\omega_p^2}{2e}(\pi-\phi) r_B. \label{Ez}
\end{equation}
Here $\phi$ describes the phase inside the wake. In this definition the rear part of the bubble corresponds to $\phi=0$ and the center is located at $\phi=\pi$. Note that the above expression is only an approximation. Especially at the rear part of the bubble the fields take often a non-linear form, depending on the electrons density distribution at this point. We will nonetheless employ equation \ref{Ez} in order to estimate the achievable energy gain in the blowout regime. The blowout radius is of the order of the spot size, so for matched conditions $k_pr_B\simeq2\sqrt{a_0}$. The maximum accelerating field in this regime is therefore of the order of
\begin{equation} 
\begin{aligned} 
E_z^{\scriptsize{\mbox{max}}} &\sim \frac{m_e c\omega_p}{e}\sqrt{a_0}\\
&\simeq96 \mbox{ GV.m}^{-1}\times \sqrt{n_0[10^{18}\mbox{ cm}^{-3}]} \times \sqrt{a_0}.
\end{aligned}
\end{equation}
The energy gain of a particle is given by
\begin{equation} 
\Delta \gamma (z) = \frac{q}{m_ec_0^2} \int_0^z E_z(z') dz'. 
\end{equation}
For a constant phase velocity $v_{\phi}$, assuming that the electron has been injected at the rear ($\phi_0=0$) and is highly relativistic from this moment on ($v_e\simeq c_0$), we can express the phase shift in the laboratory frame as
\begin{equation} 
\phi(z)=\frac{\pi}{r_B}\left( 1-\frac{v_{\phi}}{c_0} \right)z.\label{phase}
\end{equation}
We see from equation \ref{Ez} that energy gain will be only achieved until $\phi=\pi$. This is called the dephasing length
\begin{equation} 
L_d=\frac{r_B}{1-v_{\phi}/c_0}.\label{deph_length}
\end{equation}
Combining these results we find the energy gain
\begin{equation} 
\begin{aligned}  
\Delta \gamma(z) &= \frac{q}{m_ec_0^2}\times E_z^{\scriptsize{\mbox{max}}} \times \int_0^z (1-z'/L_d) dz'  \\
	&= \Delta \gamma_{\scriptsize{\mbox{max}}} \left(  2\frac{z}{L_d}-\frac{z^2}{L_d^2}  \right) 
\end{aligned} \label{gain(x)}
\end{equation}
which consists of a rapid linear acceleration in the beginning, which then saturates at $z=L_d$, see also figure 1c. The maximum gain is $
\Delta \gamma_{\scriptsize{\mbox{max}}}=E_z^{\scriptsize{\mbox{max}}}\times L_d/2. 
$
In general the gain can be maximized by lowering the plasma density of the accelerator, but as stated in the introduction this has a number of drawbacks and alternatives are desirable.

\section{Density tapered laser-wakefield accelerators}

In this section discuss the merits of density tapering in order to adapt the phase of the electron beam inside the wakefield. From equation \ref{phase} we see that the phase of electrons in the accelerator is a function of both the phase slippage $(1-v_{\phi}/c_0)$ and the cavity size $r_B$. In the preceding section we have discussed that the scaling laws of the phase velocity suggest to operate at lower plasma densities. 

\subsection{Phase-locking}

Alternatively the phase shift can be compensated by adapting the bubble radius $r_B$, ideally maintaining a phase $\phi\sim0$ to assure strongest accelerating fields. The cavity size should then change by
\begin{equation} 
\frac{dr_B}{dt} = \frac12\frac{v_e-v_{\phi}}{(1-\phi_0/2\pi)}.\label{phase_shift}
\end{equation}
The electron velocity can again be assumed $v_e\simeq c_0$, however the phase velocity of the plasma wave depends on the laser pulse evolution. In the weakly perturbative non-relativistic limit ($a\ll1$) it is reasonable to use the group velocity $v_g$, yet at higher intensities more effects become relevant, such as pulse steepening and energy depletion. While analytical models extending to the weakly relativistic regime have been proposed \cite{Benedetti:2015cz}, there exists no model for the blowout regime. It was empirically found that the pulse depletion in this system is of the order of $n_e/n_c$ lower \cite{Lu:2007eb}, which is equivalent to the etching velocity $v_{\scriptsize{\mbox{etch}}}$ in the linear regime \cite{Decker:1996ta}. In general we can approximate that the phase velocity follows a scaling of the form
\begin{equation} 
\frac{v_{\phi}}{c_0}\simeq 1-\kappa \frac{n_e}{n_c} \label{phase_velocity}
\end{equation}
with different values of $\kappa$, e.g. $\kappa\simeq0.5$ for the linear regime or $\kappa\simeq 1.5$ in the blowout regime.
The cavity size scales with the plasma wavelength, so starting from a plasma density $n_{e,0}$ the initial cavity size $r_{B,0}$ will evolve as $r_B= r_{B,0} \times\sqrt{n_{e,0}/n_e}$. Accordingly
\begin{equation}   
\frac{dr_B}{dt} = \frac12 \frac{r_{B,0}}{n_{e,0}}\left( \frac{n_{e,0}}{n_e} \right)^{3/2}\dot n_e \label{bubble_scaling} 
\end{equation}
and equations (\ref{phase_shift}-\ref{bubble_scaling}) can be combined to a first-order nonlinear ordinary differential equation of the type $\dot n_e - \alpha\times n_e^{5/2}=0$, whose solution predicts a scaling 
\begin{equation}  n_e(z) = \frac{n_0}{(1-z/L_0)^{2/3}}. \label{n(x)}   \end{equation}
The density increases first close to linearly, but then the density ramp becomes increasingly steep until a singularity is reached at $L_0=(2/3\kappa)(1-\phi/2\pi)(n_c/n_{e,0})r_{B,0}$. Using equations (\ref{deph_length}) and (\ref{phase_velocity}) we find that this length is related to the dephasing length $L_0=(2/3)(1-\phi/2\pi)L_d$.

\begin {figure}[t]
\centering
\includegraphics[width=0.8\linewidth]{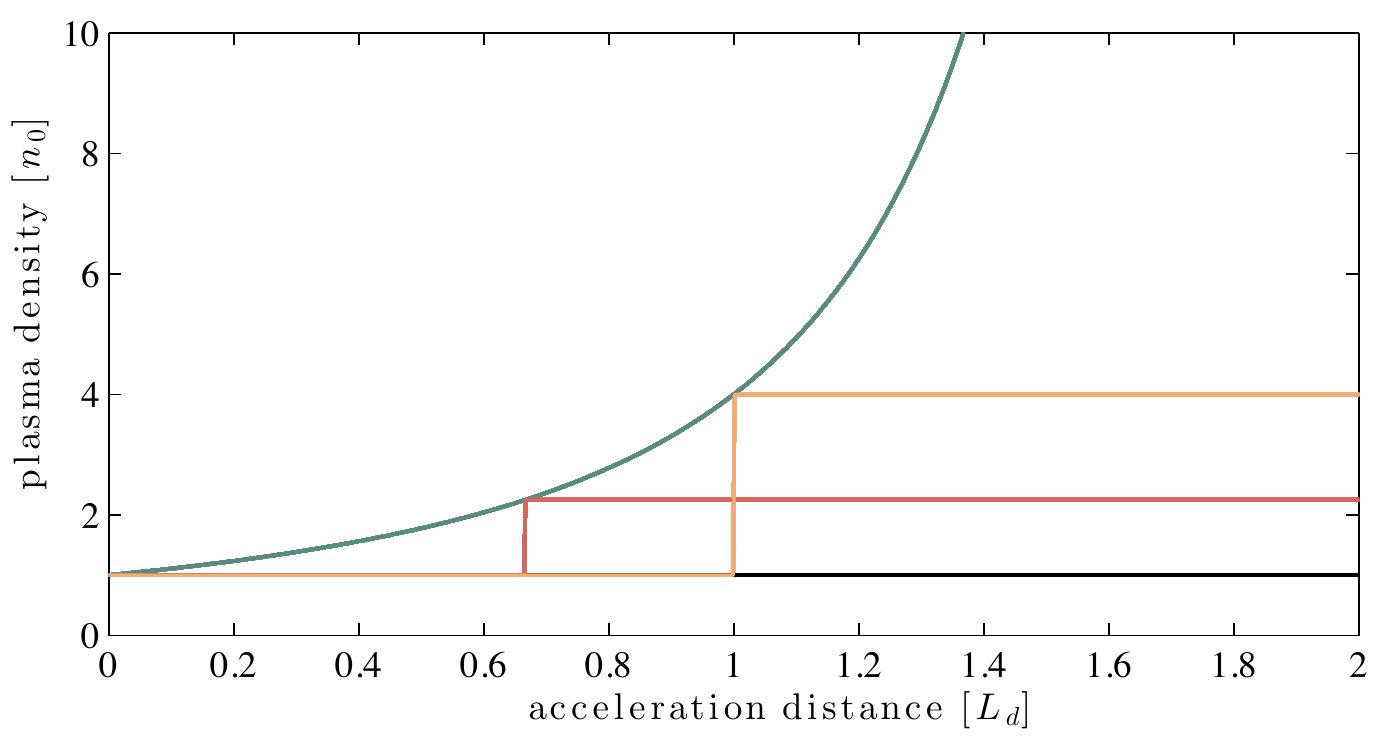}
\includegraphics[width=0.8\linewidth]{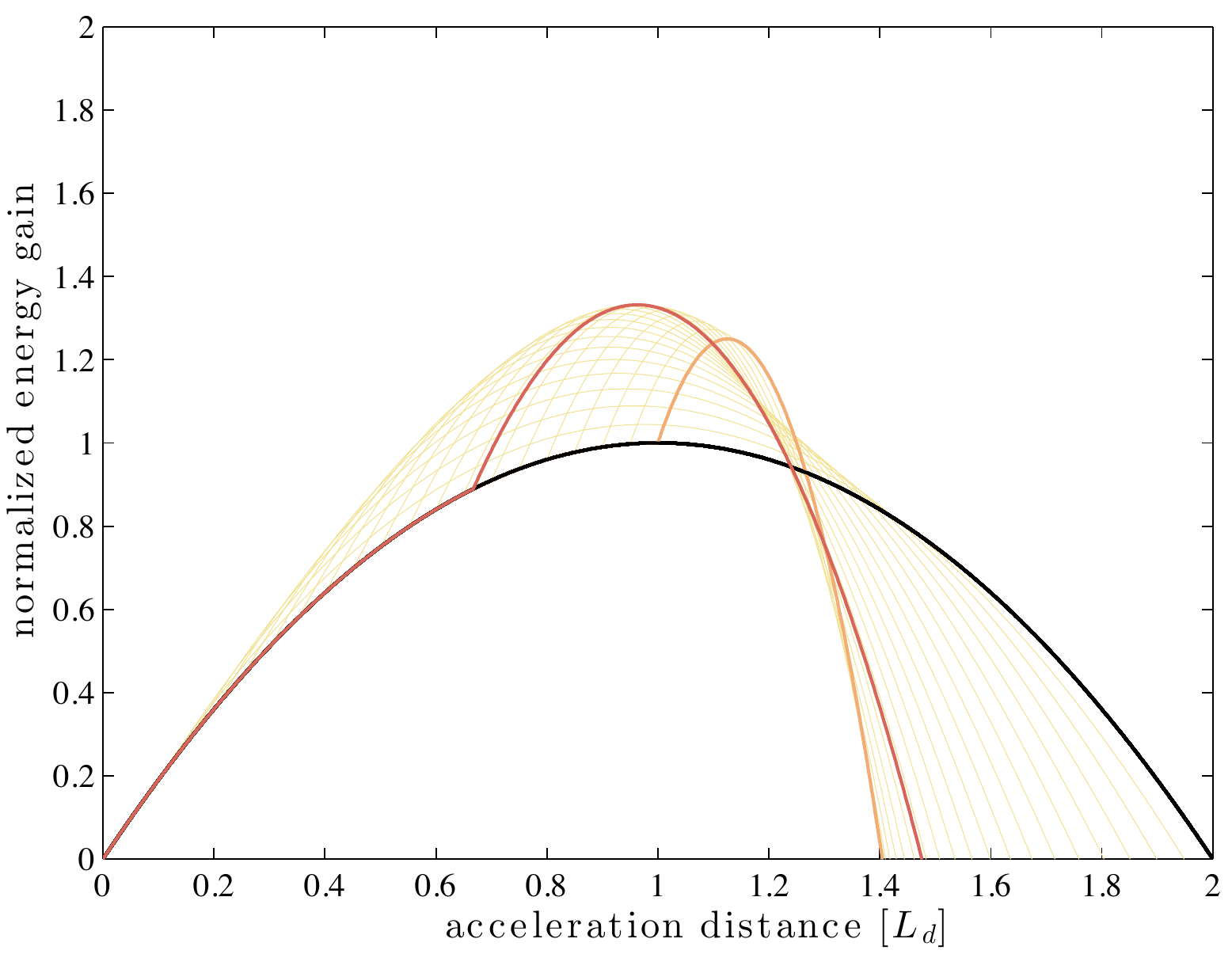}
\caption{Density profiles and estimated energy gain curve for phase-rest to $\phi=0$ at different positions along the acceleration. The density profiles for rephasing at $z=\frac23 L_d$ and $z=L_d$ are plotted again in red and orange, respectively.}
\label{model}
\end{figure}
However, in the regime of relativistic optics ($a_0>1$) the plasma wavelength scales also with the laser intensity and as mentioned before the cavity size depends also on the pulse length and width. Yet the above type of model neglects the laser pulse evolution and the coupling efficiency between the laser pulse and the plasma. For the former Sprangle and coworkers \cite{Sprangle:2001jc} have taken into account self-focusing, while Ritterhofer et al. \cite{Rittershofer:2010jg} considered the pulse evolution in plasma channels. The coupling efficiency was discussed by Pukhov and Kostyukov \cite{Pukhov:2008is}. Unfortunately all of these approaches are restricted to the linear wakefield regime ($a_0<1$) and therefore none of these descriptions are valid for the self-guided bubble regime. Here it is imperative to include ponderomotive self-focusing, relativistic self-focusing and self-compression. Increasing the plasma density amplifies these effects. In consequence the density induced change of the cavity size will be partially or completely counteracted by the augmented laser intensity. Also, the pulse depletion rapidly increases and increasing energy gain via phase-locking is therefore not straightforward if not impossible in this regime. The laser does though not react instantly to the density change, so a sharp transition as in a step-like profile might be a promising alternative.

\subsection{Phase-reset}

Instead of keeping a constant phase $\phi_0$, another conceivable situation is that an electron is first accelerated in a flat density profile and the bubble is then forced to diminuish at once, so that the particle is again in a region of accelerating fields. For a particle that has dephased to a value $\phi$, the bubble radius has to be reduced to a value $r_{B,1}=(1-\phi/2\pi)r_{B,0}$ in order to reset the phase to zero. The advantage of this scenario is that if the density transition is sharp enough, we can neglect the laser pulse evolution and assume that the cavity size is determined solely by the plasma density profile. As mentioned before, the scaling is then $r_B\propto n_e^{-1/2}$, so we find that the density $n_1$ necessary to achieve the bubble contraction for a phase reset is
\begin{equation}  
n_1(\phi)=\frac{n_0}{(1-\phi/2\pi)^2}.
\label{n_1(z)}
\end{equation}  

We can now calculate the energy gain for such a phase reset in the blowout regime. For a first estimation we assume a boost when the electron is just dephased, i.e. $\phi=\pi$ or equivalently $z=L_d$. In this case the required density transition (\ref{n_1(z)}) is $n_1=4n_0$. Once rephased, the electrons will essentially behave as if they were just injected into a new accelerator with density $n_1=4n_0$. So we can use (\ref{gain(x)}) and sum the dephasing limited gain of those two 'stages', which gives
\begin{equation}  \Delta \gamma_{max} =  \Delta \gamma_{max} (n_0) +  \Delta \gamma_{max} (4n_0) = \frac54 \Delta \gamma_{max} (n_0).\end{equation}
The second stage contributes much less to the overall gain, as the dephasing length is shorter at higher density. Still, we expect a gain of around 25 percent.

\begin {figure}[t]
\centering
\includegraphics[width=1.\linewidth]{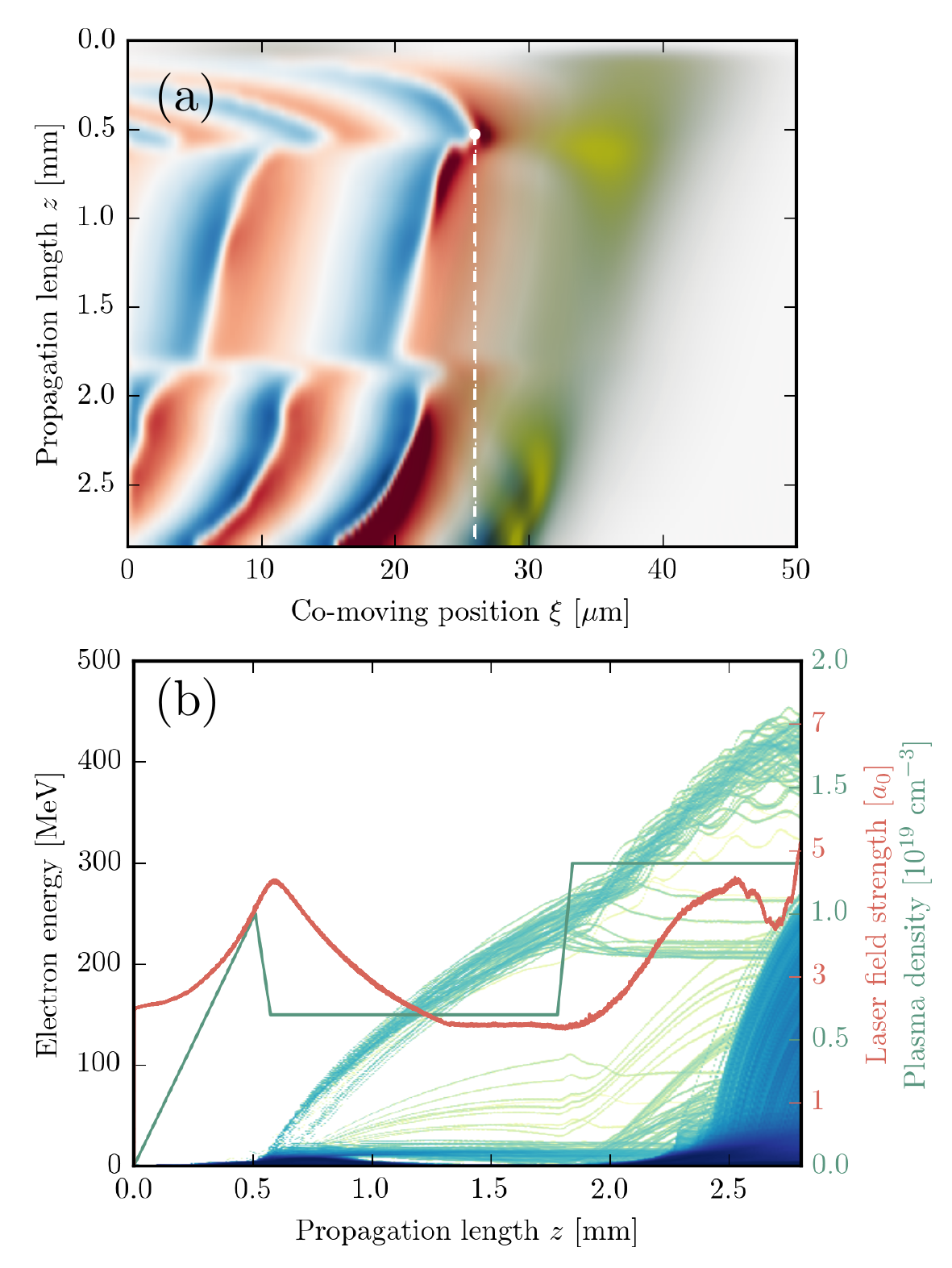}
\caption{PIC simulation of electron acceleration with rephasing in a density step. The total gain is about 30 percent higher than for the untapered profile.}
\label{rephasing1}
\end{figure}

Let us now evaluate the optimal case. To find the maximum achievable energy gain in a unique boost we take the gain in a sawtooth-shaped wakefield (eq. \ref{gain(x)}) up to a position $z_{\scriptsize{\mbox{boost}}}$
\begin{equation} \Delta\gamma_1= \left(  \frac{2z_{\scriptsize{\mbox{boost}}}}{L_d} - \frac{z_{\scriptsize{\mbox{boost}}}^2}{L_d^2} \right) \times  \Delta \gamma_{max} (n_0)  \end{equation}
and add the density dependent gain according to the rephasing density (\ref{n_1(z)})
\begin{equation} \Delta\gamma_2=\left(1-\frac{z_{\scriptsize{\mbox{boost}}}}{2L_d}\right)^2 \times  \Delta \gamma_{max} (n_0). \end{equation}
The complete energy gain is therefore
\begin{equation}  
\begin{aligned} 
	\Delta\gamma=  \left( 1+\frac{z_{\scriptsize{\mbox{boost}}}}{L_d}-\frac34\frac{z_{\scriptsize{\mbox{boost}}}^2}{L_d^2}\right) \times  \Delta \gamma_{max} (n_0).
	\label{3_boost_gain}
\end{aligned}
\end{equation}
As shown in figure \ref{model}, the global maximum is located before the actual dephasing at $z=2/3\times L_d$. It leads to $\Delta \gamma_{max}\left(\frac23L_d\right) =  \frac43 \Delta \gamma_{max} (n_0)$. So we estimate that a phase reset can lead to gain in the order of one third of the dephasing limited energy gain. The results are in accordance with \cite{Phuoc:2008ck}, where such a density step scenario was empirically studied using test particle simulations. It is worth noting that this relative energy augmentation is independent of the plasma density $n_0$. However, this situation will change if we take into account the laser pulse evolution. As we will discuss in the next section, the scheme is most likely to work best at high densities, where electrons gain significant energy over short distances. Furthermore, nonlinearities in the wakefield, which occur especially at the back of the bubble, may also increase the actual gain of the scheme.

\subsection{Particle-in-cell simulations}

In order to validate the predictions from the phenomenological model presented in the preceding section, we have performed three-dimensional particle-in-cell simulations using the quasi-cylindrical code \textsc{Calder-Circ} \cite{Lifschitz20091803}. We use two azimuthal modes (m $= 0 - 1$) and a $1500\times 250$ mesh in the longitudinal and radial directions, respectively. The numerical resolution is $\Delta z=0.3k_0^{-1}$, $\Delta r=1.5k_0^{-1}$ and $c_0\Delta t=0.96\Delta z$ (with $k_0 =1/\lambda_0$). The simulation box moves forward at the speed of light in vacuum, so the coordinates $\xi=z-c_0t$ are almost co-propagating with the laser. Yet the laser pulse is still slowly dephasing with both electrons (moving also at a velocity close to $c_0$) and the simulation box. This can be seen for instance in figure \ref{fig1}(b), where it is apparent that the laser pulse (yellow) moves backwards in the simulation box.

The laser pulse is modeled similar to the parameters of the \textsc{Salle Jaune} laser at Laboratoire d'Optique Appliqu\'ee, with a duration of 30 femtoseconds, a waist of 11.5 micrometers and a peak intensity $a_0=2.5$. In order to avoid beam-loading effects, which also alter the cavity size, it is preferential to operate with a weakly charged electron beam. Our reference simulation uses a 50 micrometer density transition from $n_e=1\times10^{19}$ cm$^{-3}$ to $0.6\times10^{19}$ cm$^{-3}$ to inject a well-localized electron beam ($\sigma_z<1\mu$m) into the laser wakefield \cite{Suk:2001uo}. Following injection the density of the reference case remains constant. This density of the plasma is chosen relatively low ($n_e<10^{19}$ cm$^{-3}$) in order to avoid self-injection and there is an initial density upramp which prevents injection before the transition. The results of this simulation are shown in figure \ref{fig1}. We observe that the electrons reach a maximum energy of about 300 MeV after 2 millimeters of acceleration, corresponding to an average field gradient of $\sim150$ GV/m.  

For the rephasing case we boost the beam after $z\simeq0.6L_d$, which according to equation (\ref{n_1(z)}) requires a density increase from $n_0$ to $2n_0$. The transition length is $50\mu$m. As shown in figure \ref{rephasing1} the density step steepens the energy gain curve, resulting in a final beam energy of around 400 MeV. This 30 percent increase is very similar to the predictions from the model, see figure \ref{model}. However, as we have discussed in the preceding sections, after some hundred microns of propagation self-focusing sets in, which then leads to self-injection. The large amounts of electrons injected through this mechanism provoke a cavity expansion which brings the shock injected beam faster into dephasing. We also observe that a small part of the beam is lost during rephasing. This is surprising, as we see in figure \ref{rephasing1}a that the electron beam (white dashed line) does not reach the rear of the bubble during the rephasing. However, maintaining electron at the rear part of the bubble is delicate, because the focusing fields are weak in this region. 

\begin {figure}[t]
\centering
\includegraphics[width=1.\linewidth]{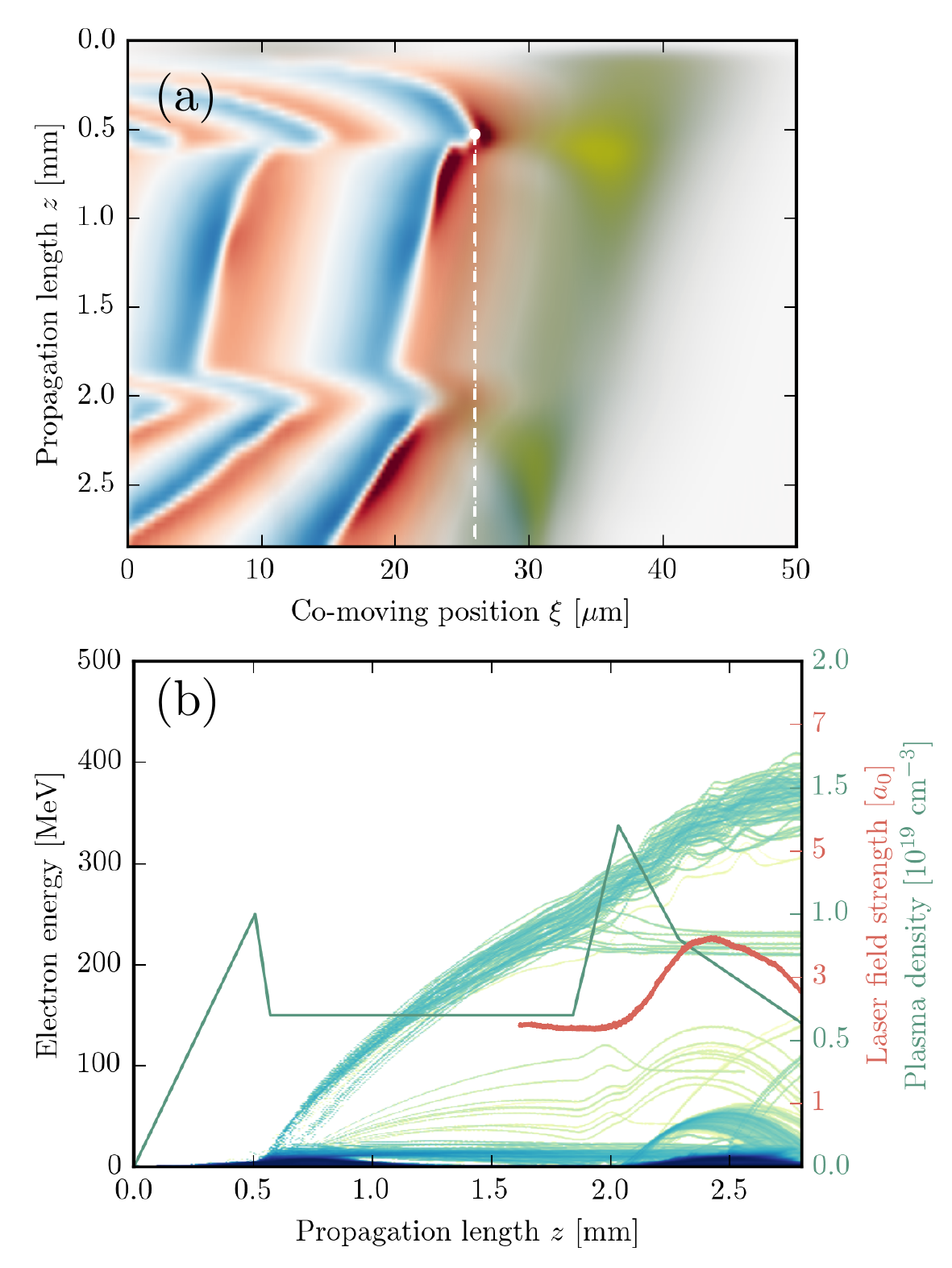}
\caption{PIC simulation of electron acceleration with rephasing in a shock-like density profile. The energy gain has reduced to about 20 percent, but in turn self-injection is suppressed.}
\label{rephasing2}
\end{figure}

In general we observe that the self-focused laser field strength $a_0$ evolves delayed, but almost linear with the density profile. In consequence all cavity contractions in this regime are eventually compensated and if the density is high enough self-injection is triggered. We have therefore tested another profile, which reduces the density again after the step. As shown in Fig.\ref{rephasing2}, the laser pulse evolves much weaker and self-injection is suppressed. The energy gain is reduced in this configuration, yet the gain is still 20 percent in this simulation and can be further optimized.

\begin {figure}[t]
\centering
\includegraphics[width=0.8\linewidth]{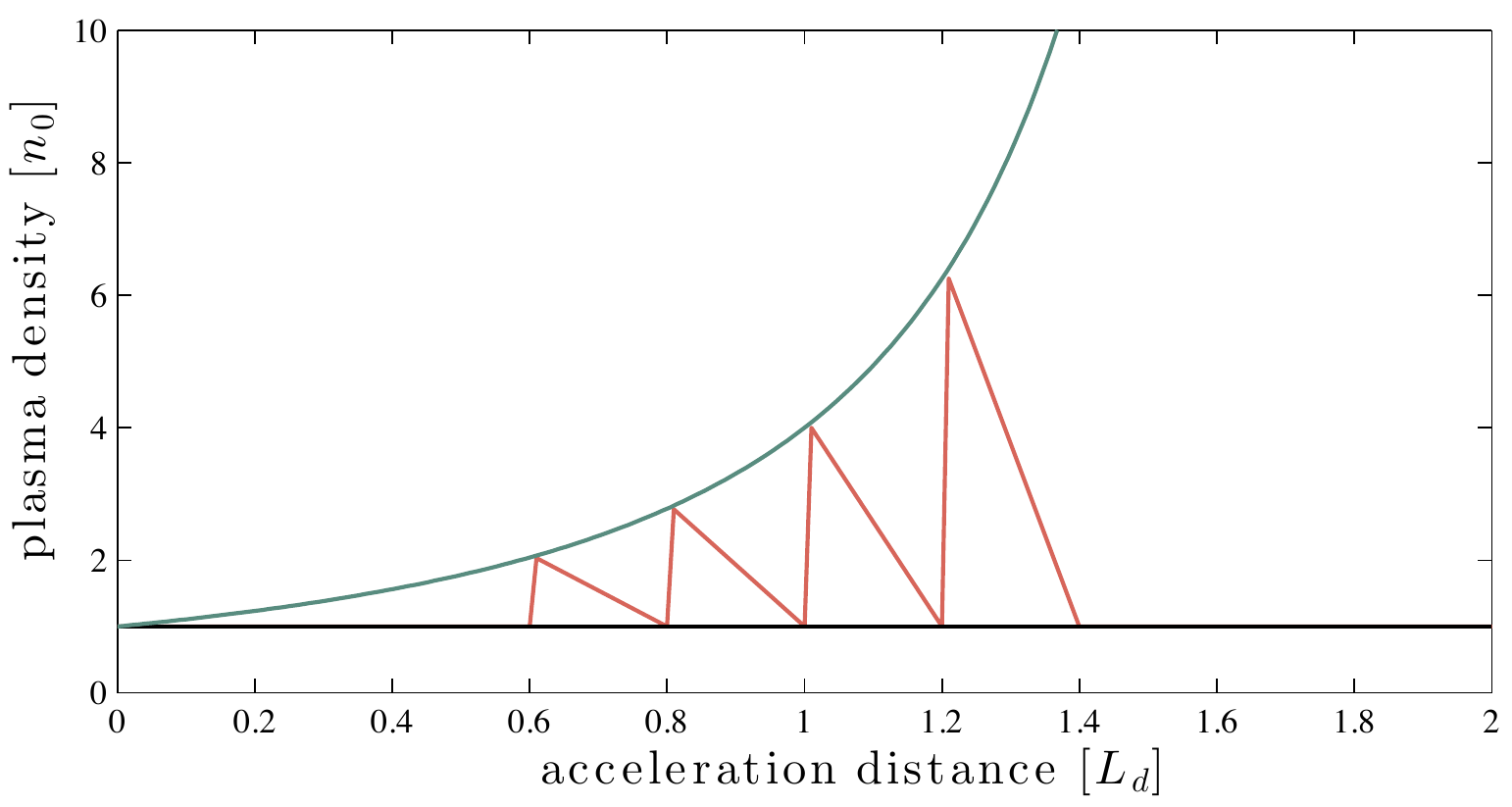}
\includegraphics[width=0.8\linewidth]{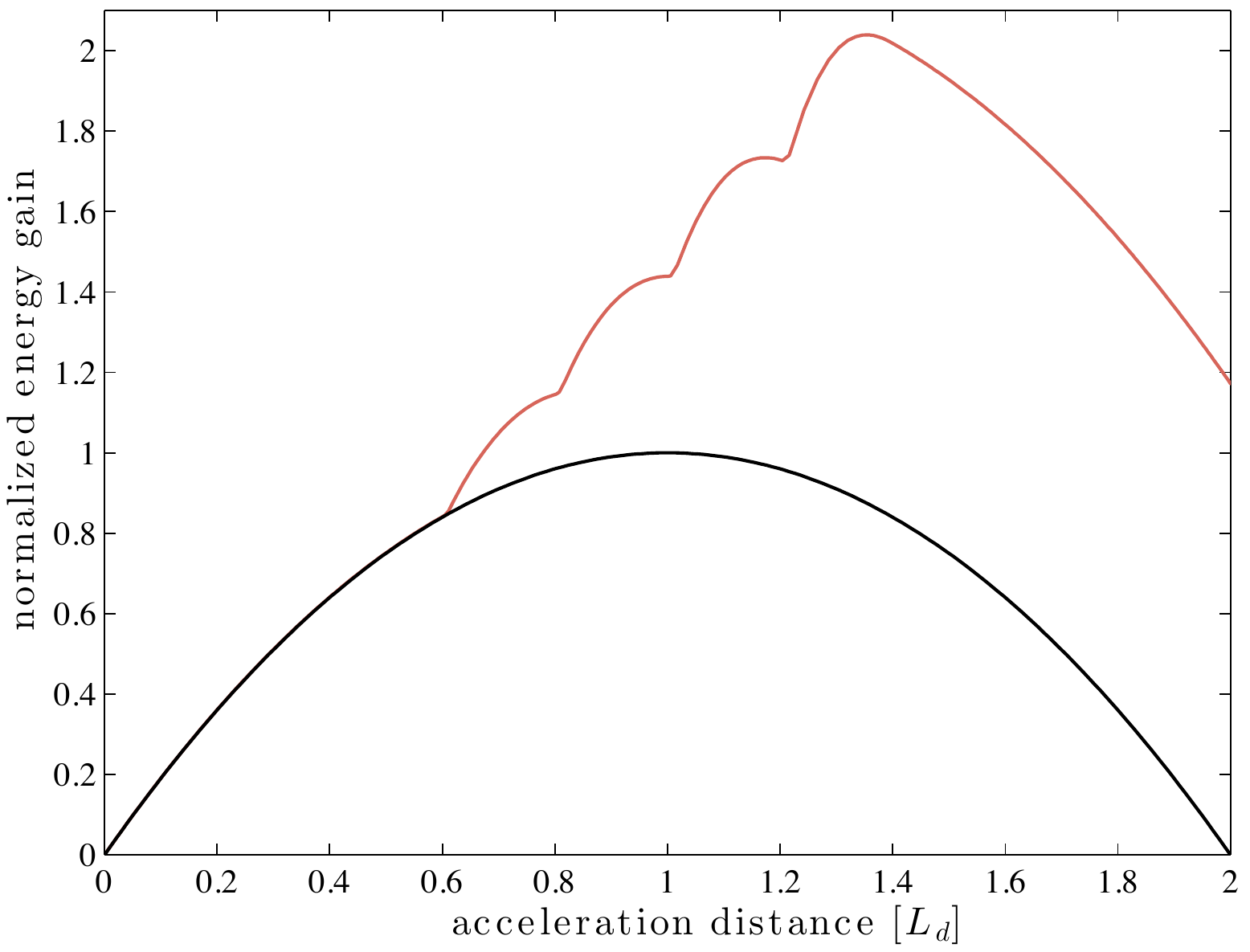}
\caption{Estimation for energy gain via successive rephasing in sawtooth-shaped density transitions.}
\label{stages}
\end{figure}

\subsection{Multiple rephasing stages}

The question arrises how much energy gain is achievable in a phase-reset scheme. In the preceding sections we have seen that the density profile should be tailored in a way that reduces the laser pulse evolution \cite{Mourou:2006ef}, which is essentially relativistic self-focusing, ponderomotive self-focusing and self-compression.

As an example we have calculated the gain that would be achievable in a set of consecutive rephasing stages, each having a sawtooth-like density profile. As shown in figure \ref{stages} it might be possible to achieve more than a twofold increase in the final beam energy in such a configuration. However, for these calculations we assumed that the laser pulse does not evolve significantly in this profile, which has to be confirmed in particle-in-cell simulations.

It is also important to point out the differences of rephasing to other concepts of staging in laser-wakefield accelerators \cite{Malka:2006ff}. To date only single-pulse staging schemes have been experimentally realized \cite{Pollock:2011hf,Liu:2011ex,Kim:2013dc} and none of these schemes does influence the phase between electrons and laser. So the maximum distance between driver and witness is fixed during injection, which is why the energy gain in such accelerators is basically limited by the injector stage. Increased energy gain would be achievable in multi-pulse setups, where the phase can be reset by use of a new laser pulse that can be synchronized independently of the electron bunch. However, such a setup is much more difficult to realize experimentally since it requires very good alignment and synchronization of the laser and particle beam. Also, additional beam optics such as plasma lenses \cite{Thaury:2015cg,vanTilborg:2015eo} are required to maintain the electron beam emittance during transport from one stage to the next one. \\

\section{Conclusions}

In conclusion we have studied the possibility of increased electron energy gain using density tapering. In contrast to preceding studies we focused on the phase reset in sharp density transitions. Assuming linear fields inside the blowout region we estimate that a unique phase reset can lead to a gain in the order of 30 percent. Similar values are reproduced in particle-in-cell simulations. As seen in \cite{Guillaume:2015di} even higher gains are achievable when the fields become non-linear. 
Furthermore we have discussed the gain in sawtooth-shaped density profiles which suppresses the laser pulse response to the high plasma densities. We estimate that several stages of such rephasing can lead to a twofold increase in beam energy, which has to be confirmed by comprehensive simulations.\\

ACKNOWLEDGMENTS: We acknowledge the Agence Nationale pour la Recherche through the FEN- ICS Project No. ANR-12-JS04-0004-01, the Agence Nationale pour la Recherche through the FEMTOMAT Project No. ANR-13-BS04-0002, the X-Five ERC project (Contract
No. 339128), EuCARD2/ANAC2 EC FP7 project (Contract No. 312453), LA3NET project (GA-ITN-2011-289191), and GARC project 15-03118S.


\end{document}